\documentclass[prb,aps,twocolumn,showpacs,amsmath,amssymb]{revtex4-1}
\usepackage{float}
\usepackage{amsmath}
\usepackage[dvips,final]{graphicx}
\usepackage{epsfig}
\usepackage{color}

\begin{document}

\title{Interaction potential of FePt with the MgO(001) surface}

\author{R. Cuadrado}
\affiliation{Department of Physics, University of York, York YO10 5DD,
             United Kingdom}

\author{R. W. Chantrell}
\affiliation{Department of Physics, University of York, York YO10 5DD,
             United Kingdom}

\date{\today}

\begin{abstract}
By means of density functional theory we have undertaken a structural, electronic
and magnetic survey of the adsorption of the Fe$_x$Pt$_y$~($x,y\leq 4$) clusters 
on MgO(001) surface under the generalized gradient approximation. We have tested 
different atomic adsorption geometries with the aim of scan a wider range of adsorption 
sites in order to determine the preferential surface covering. Our main conclusion 
in this respect is that the FePt wets the surface. The $intracluster$ (before and 
after the adsorption) and cluster--to--surface binding mechanisms were investigated 
via the adsorption energy, charge transfer, density of states and hybridization 
analysis. The adsorption energy values increased for those geometries in which 
keeping the Fe or Pt atom @$top$--O, the outermost species was moved to $cover$ 
the surface. In general the unsupported clusters present higher $intracluster$ energies 
than the adsorbed ones being the average difference of 1.5~eV. In this regard there 
was a small reduction in the net magnetic moment of the supported clusters due to 
an internal and external rearrangement of the spin--up/--down charge. Furthermore, 
a complex and subtle charge transfer between different species takes place having 
an increasing the Pt and O population at the expense of the lost Fe charge. 

\end{abstract}


\maketitle

\section{Introduction}\label{intro-sec}
During the past two decades, continuous miniaturization has driven 
data-storage technology to the nanometer scale and consequently, the study 
of magnetism in low-dimensional systems is currently attracting a great 
deal of interest~\cite{shou,hans}. The interaction of metallic clusters and
alloys with supporting metal–-oxide surfaces is a subject of great current 
interest, because of their numerous technological applications. 
\cite{musolino,ricci} Important objectives of these studies are to understand 
how the atomic and electronic structure of both subsystems are modified through 
their interaction, as well as the properties of the resulting interface. Due to 
its high uniaxial magnetocrystalline anisotropy energy~(MAE) of $7\times 10^7$ 
ergs/cm$^3$, the FePt--L1$_0$ alloy is a promising candidate for the next 
generation for the fabrication of ultrahigh density data-recording devices. 
FePt in the disordered state has a face--centered cubic ($fcc$) crystal 
structure and is magnetically soft. In contrast, the ordered face--centered 
tetragonal~($fct$) L1$_0$ phase of the FePt alloy has Fe and Pt atomic planes 
stacked alternatively along the $c$ axis in the equiatomic composition. In 
this phase it exhibits exceptional magnetic properties including high 
uniaxial~(MAE), high coercivity, high saturation magnetization, and good 
chemical stability.~\cite{weller} There have been a number of studies 
looking at the magnetic properties of FePt--L1$_0$, for example, in films or 
multilayers,~\cite{thin,multi} slabs~\cite{Chepulskii2012} and more recently 
in doped bulk phases.~\cite{ramon2} 

However, during the last decade the possibility 
to use clusters deposited on surfaces to increase the recording density.~\cite{dieguez,
cuadrado1,gambardella,medina} has emerged. These clusters or nanoparticles~(NPs) have 
different properties from bulk counterparts due to their reduced surface 
atomic coordination. In particular, binary 3$d$--5$d$ NPs formed by transition 
metals~(TM) such as Fe or Co, together with 5$d$ noble metals such as Au or 
Pt allow the possibility to tune the magnetic properties based on an in--depth 
knowledge of their geometrical and magnetic behavior.~\cite{rollman,seivane,chen} 
The high MAE also permits the use of smaller particles before the onset of 
superparamagnetism, which may translate into potentially larger recording 
densities. Hence, the primary motivation to synthesize and study perpendicularly 
oriented FePt nanoparticles dispersed in a nonmagnetic matrix is to combine 
favorable properties, which accompany size reduction, with superior magnetic 
properties of FePt and arrive at a system that is highly suitable for 
high-density perpendicular magnetic recording.

Wettability and material spreading are of key importance for many applications. 
At large scales, wetting or non--wetting plays an important role in, for example, 
oil recovery~\cite{bertrand} and on a smaller scale, wetting solutions have been 
proposed to solve technological problems in microfluidics and nanoprinting, 
inkjet printing, etc.~\cite{tabeling} All these phenomena are governed by the 
surface and interfacial interactions, acting usually at small~(a few nanometers 
for van der Waals or electronic interactions) or very small molecular 
distances. These length scales are now being probed with relatively new 
experimental techniques, such as atomic force microscopy, or theoretical tools, 
such as molecular dynamics.~\cite{musolino} In such surface--anchored cluster/alloy 
systems it would be highly desirable to find ways to control and tune the 
properties of the adsorbed clusters through manipulation of the supporting 
substrate and the deposited clusters. The properties that may be influenced via 
substrate manipulation include: adsorption energies, cluster geometries and 
dimensionalities, cluster diffusion barriers, charge distributions, and 
chemical reactivities.

The magnetic properties of gas--phase metal clusters have been the subject 
of both experimental and theoretical investigations  as we have pointed out, 
however, to the best of our knowledge there have been no studies about the 
wettability of the FePt--L1$_0$ alloy onto an MgO surface. The growth of 
FePt on MgO has been predicted to be one of the possible candidates 
for for the fabrication of ultrahigh density data recording media. Thus, in 
this study, we have performed detailed density functional theory~(DFT) 
computations to study the adsorption of bimetallic Fe$_x$Pt$_y$ clusters on 
MgO, where $x+y\leq 4$. We present the study of the structural, electronic and 
magnetic properties of these ultrasmall clusters regarding different adsorption 
sites for the Fe or Pt species, namely, @$top$--O/--Mg and @$hollow$ sites. Even 
though the size of these small agregates is far from the real structures currently 
used as a recording magnetic media, the look into the complicated bonding mechanism 
at the atomic level between Fe or Pt species with the MgO surface will elucidate 
the still unclear interplay between these ferromagnetic--oxide systems. The 
importance of how FePt--L1$_0$ grows onto MgO is of vital importance 
in magnetism because of the following questions: 
(1) what are the stable configurations for the adsorption of these Fe$_x$Pt$_y$ 
clusters on MgO? (2) How do the electronic and magnetic properties of the adsorbed 
clusters vary with the sizes? (3) How does the chemical activity of these 
functionalized MgO by Fe$_x$Pt$_y$ clusters?

The paper is structured as follows. In section~\ref{tools-sec} we describe the 
theoretical tools to perform all the calculations as well as the set of geometries 
studied in the present work. Convergence tests are presented in Sec.~\ref{test-subsec}. 
The energetic and structural analysis after the relaxation of the $dimers$, 
$trimers$ and $tetramers$ will be explained in~\ref{structure-subsec}. The 
$intracluster$ and the cluster--to--surface charge transfer study are described in 
subsection~\ref{dos-sec} together with the local magnetic moments analysis. Finally, 
Sec.~\ref{conclusion-sec} summarizes the main results.

\section{Theoretical Methods}\label{tools-sec}
Our density functional based calculations have been performed using the code 
SIESTA~\cite{siesta} within the generalized gradient approximation~(GGA) for 
the exchange correlation~(XC) potential~\cite{pbe}. We used norm-conserving
pseudopotentials in the separate Kleinman-Bylander~\cite{kb} form under the
Troullier-Martins parametrization~\cite{tm} and to address a better description
of the magnetic behaviour, nonlinear corrections were included in the XC 
terms~\cite{cc}. The valence reference electronic configuration for the 
pseudopotentials are 4$s^2$4$p^0$3$d^6$~(Fe), 6$s^1$6$p^0$5$d^9$~(Pt), 
3$s^2$3$p^0$3$d^0$~(Mg) and 2$s^2$2$p^0$3$s^2$~(O) with $s/p/d$ cutoff 
radii 2.0/2.5/0.6 a.u.~(Fe), 2.0/2.75/1.25 a.u.~(Pt), 2.6/2.6/2.6 a.u.~(Mg) 
and 0.75/0.75/1.75 a.u.~(O), respectively. The geometry optimizations were 
carried out using the conjugate gradient~(CG) method at a spin-polarized 
scalar relativistic level. As a basis set, we have employed double-$\zeta$ 
polarized~(DZP) with strictly localized numerical atomic orbitals, and, the 
electronic temperature --kT in the Fermi-Dirac distribution-- was set to 
50~meV. After the relaxation process the forces per atom were less than 
3~meV/\AA.   

The adsorption of the Fe$_x$Pt$_y$ clusters onto a MgO surface has been 
modeled as a two--dimensional periodic slab comprising of four MgO(001) 
layers. We first consider $dimer$ structures as illustrated schematically 
in Fig.~\ref{dimer-fig}, followed by more complex FePt clusters as depicted 
in Fig.~\ref{tr+te-fig}. The Magnesium oxide structure can be 
described as two interpenetrating {\it fcc} lattices displaced by $a/2(111)$ 
along the body diagonal of the conventional cube and the bulk experimental 
value for its lattice parameter is $a$=4.22\AA. We optimized the MgO lattice 
constant $a$ for the GGA--XC functional obtaining the lattice value of 4.30\AA. 
To converge the physical quantities and minimize the out--of--plane and 
in--plane interactions between nearest cells, a total of subtract32 in--plane 
MgO atoms was chosen, as we will point out in the section~\ref{test-subsec}. 

In this work, the adsorption energies, E$_{ads}$, were evaluated after 
subtracting from the total energy of each configuration of the energy 
of the clean MgO surface and those of the clusters following the equation:
\[
E_{ads} = -(E^T - E_C^T-E^{at/clus}),
\] 
where $E^T$ is the total energy of the whole system, the $E_C^T$ is the 
energy of the clean MgO surface and the $E^{at/clus}$ is the energy if 
either one Fe/Pt atom or Fe$_x$Pt$_y$ clusters are adsorbed. 

\section{Results and discussion}\label{results-sec}
We performed the energetic study of different size FePt clusters adsorption 
on MgO(001) surface. Several adsorption sites and clusters orientations with 
respect to the MgO surface were taken into account. The optimized geometric 
configurations were obtained by means of the conjugate gradient method allowing 
that the clusters and the first two MgO layers moved freely. In the following 
we present the sistematic analysis of the geometric, electronic and magnetic 
properties of these optimized configurations checking previously the feasibility 
of the calculations with a convergency test for two unit cell sizes. 

\subsection{Convergence test}\label{test-subsec}

\begin{table}
\caption{Adsorption energies, E$_{ads}$, heights of the Fe/Pt adatoms with respect 
         to the first MgO plane, $z_{Fe/Pt}$, for the $top$--O/--Mg and $hollow$  
         adsorption sites for different number of support MgO planes, $N_l$, 
         and for two sizes of the unit cell: c18 and c32. The energies are in eV 
         and the heights in \AA. 
         \label{ads-energies}}
\begin{tabular}{ccccccccccc} 
&&&&&&&&&& \\ \hline \hline
&&&\multicolumn{3}{c}{{\bf Fe adatom}}&&&\multicolumn{3}{c}{{\bf Pt adatom}}\\ \hline  
   cell    &Fe/Pt site & N$_l$ &E$_{ads}$&&z$_{Fe}$  &&&E$_{ads}$&&z$_{Pt}$ \\ \hline
 {\bf c18} &  $top$--O  &  2 &  2.26   &&  1.99    &&&  3.54   &&  2.00    \\
           &           &  3 &  2.10   &&  1.92    &&&  3.64   &&  2.00    \\
           &           &  4 &  2.25   &&  1.93    &&&  3.50   &&  2.00    \\
\vspace{0.05cm}
           &           &  5 &  2.32   &&  1.97    &&&  3.35   &&  2.00    \\
           & $top$--Mg  &  2 &  0.69   &&  2.87    &&&  1.63   &&  2.64    \\
           &           &  3 &  0.87   &&  2.70    &&&  1.44   &&  2.60    \\
           &           &  4 &  0.71   &&  2.73    &&&  1.91   &&  2.65    \\
\vspace{0.05cm}
           &           &  5 &  0.67   &&  2.75    &&&  1.16   &&  2.64    \\
           &$hollow$   &  2 &  1.74   &&  1.96    &&&  2.67   &&  2.05    \\
           &           &  3 &  1.88   &&  1.90    &&&  2.70   &&  2.03    \\
           &           &  4 &  1.56   &&  1.96    &&&  2.65   &&  1.98    \\
\vspace{0.2cm}
           &           &  5 &  1.50   &&  1.95    &&&  2.71   &&  1.98    \\
 {\bf c32} & $top$--O   &    2   &  2.40   &&  1.93    &&&  3.51   &&  2.00     \\
           &           &    3   &  2.17   &&  1.92    &&&  3.56   &&  2.01     \\
           &           &    4   &  2.29   &&  1.93    &&&  3.62   &&  2.00     \\
\vspace{0.05cm}
           &           &    5   &  2.28   &&  1.93    &&&  3.70   &&  2.01     \\
           & $top$--Mg  &   2   &  0.84   &&  2.97    &&&  1.35   &&  2.57     \\
           &           &   3   &  0.87   &&  2.92    &&&  1.86   &&  2.67     \\
           &           &   4   &  0.88   &&  2.94    &&&  1.53   &&  2.64     \\
\vspace{0.05cm}
           &           &   5   &  0.88   &&  2.91    &&&  1.53   &&  2.66     \\
           & $hollow$  &   2   &  1.26   &&  1.95    &&&  2.70   &&  1.99     \\
           &           &   3   &  1.75   &&  1.94    &&&  2.72   &&  1.95     \\
           &           &   4   &  1.64   &&  1.96    &&&  2.80   &&  1.99     \\
           &           &   5   &  1.49   &&  1.95    &&&  2.53   &&  2.00     \\ \hline \hline
\vspace{0.05cm}
\end{tabular}
\end{table}

In order to use the best geometric configuration to avoid the spurious 
interaction between adjacent in--plane and out--of--plane cells due to the 
periodic boundary conditions, we have calculated the adsorption energy of 
one Fe and Pt atom on the @$top$--O, @$top$--Mg and @$hollow$ sites as well 
as the vertical distances between these atoms and the MgO surface. In order 
to ensure that our model systems were sufficiently large we have investigated 
the effect of increasing the MgO system size. Specifically, we have chosen 
two supercell sizes labeled as $c18$ and $c32$ with 18 and 32 MgO atoms at 
the surface respectively, and varied the thickness of the supported MgO 
layer by varying the number of atomic planes from 2 to 5. Each configuration 
was optimized using the conjugate gradient~(CG) method until the forces 
between atoms were less than 0.03~eV/\AA. The relaxations were undertaken 
allowing the Fe or Pt atoms and the first two MgO planes move freely, keeping 
the atoms belonging to the last MgO layers fixed to their bulk sites. Only for 
the configurations in which the number of MgO planes were two or three, i.e., 
$c18/32$--2/3, only the first MgO layer was allowed to move together with the 
adsorbed atoms. In table~\ref{ads-energies} we can see the values of the 
adsorption energies~(E$_{ads}$) and the heights between the Fe/Pt atoms and 
the surface~($z_{Fe/Pt}$). These values were computed as the difference in 
the $z$ coordinate between the Fe/Pt and the O/Mg species, depending on whether 
the adsorption site was @$top$--O or @$top$--Mg, respectively. The perpendicular 
distances for atoms situated on @$hollow$ sites were calculated as the difference 
of the $z$ coordinate of the Fe/Pt and the  average $z$ coordinate of O and Mg. 

Detailed analysis of table~\ref{ads-energies} is not germane to the aim of the 
paper, so here we only discuss the main results in order to discriminate the 
less accurate geometric configurations and define the best geometry for accurate 
and CPU efficient calculations of the properties of FePt clusters on MgO. We expect 
that a bigger supercell together with a higher number of MgO layers will provide 
more accurate calculations. The reason is clear after inspection of the differences 
in the E$_{ads}$ as we move from two to five MgO layers and after we compare 
the values between the $c18$ and $c32$ supercells. As an example, in 
$c18$--Fe@$top$--O configurations the E$_{ads}$ converges to 2.32~eV 
as the MgO thickness increases. Almost the same value is achieved when the 
$c32$ was used~(2.28~eV). This suggests that the use of the $c32$ 
configuration composed of 4 or 5 MgO layers is sufficient to study 
the adsorption of the FePt clusters onto MgO(001) surface. Beyond this there 
are no significant changes in the distances between the atoms and the surfaces, 
only around $\pm$0.05~\AA, in moving to the higher thickness and sizes. 

It is interesting to note that the Pt atoms prefer to lie @$top$--O as shown 
in the adsorption energies in comparison with the other adsorption 
positions and species in the table~\ref{ads-energies}. This is surprising 
because, as shown in~\cite{FePtOnMgO}, the Fe atoms 
of a FePt--L1$_0$ alloy prefer to lie @$top$--O rather than having a
Pt-termination. Clearly the behavior of individual Fe and Pt atoms is very 
different from that of a FePt alloy on MgO(001). However, we will show later 
that as the size of the FePt clusters increases the preferential adsorption 
of the Fe species returns to the @$top$--O site after adding one Pt atom to 
the cluster.
   
\subsection{DFT structural relaxations}\label{structure-subsec}
The main purpose of this work is to shed light on whether 
the magnetic FePt--L1$_0$ alloy prefers to grow vertically or 
horizontally when it is supported by the MgO(001) surface. If the 
magnetic alloy grows, covering the whole surface, it is said that the 
FePt wets the MgO surface. In the other case the L1$_0$--like structure 
will cover remain localised to region on the MgO surface forming FePt--L1$_0$ 
``drops''. The most obvious means of determining whether the FePt wets the 
MgO would be to simulate large extended structures to determine the preferential 
growth. However, such calculations would be extremely computationally 
expensive and beyond the scope of this paper. Instead, we have used only a 
representative number of ultrasmall Fe$_x$Pt$_y$ clusters with different 
sizes and structures, ranging $x$ and $y$ from 1 to 2. As we will see, 
we will fix some atoms of the clusters @$top$--O on the MgO surface and 
will move others in order to scan different adsorption positions, keeping 
the same number of the total atoms in the cluster. As we pointed out in 
section~\ref{test-subsec} we have used four MgO planes and the $c32$ 
supercell for all the calculations. The Fe$_x$Pt$_y$ clusters and the 
first two MgO planes were allowed to move freely during the optimization, 
keeping the atoms of the last two planes fixed to their bulk positions. 
The final forces on the atoms involved were less than 0.03~eV/\AA.     

\subsubsection{Dimers on MgO}\label{dimers-subsec}
\begin{figure}[tb]
 \includegraphics[scale=0.30]{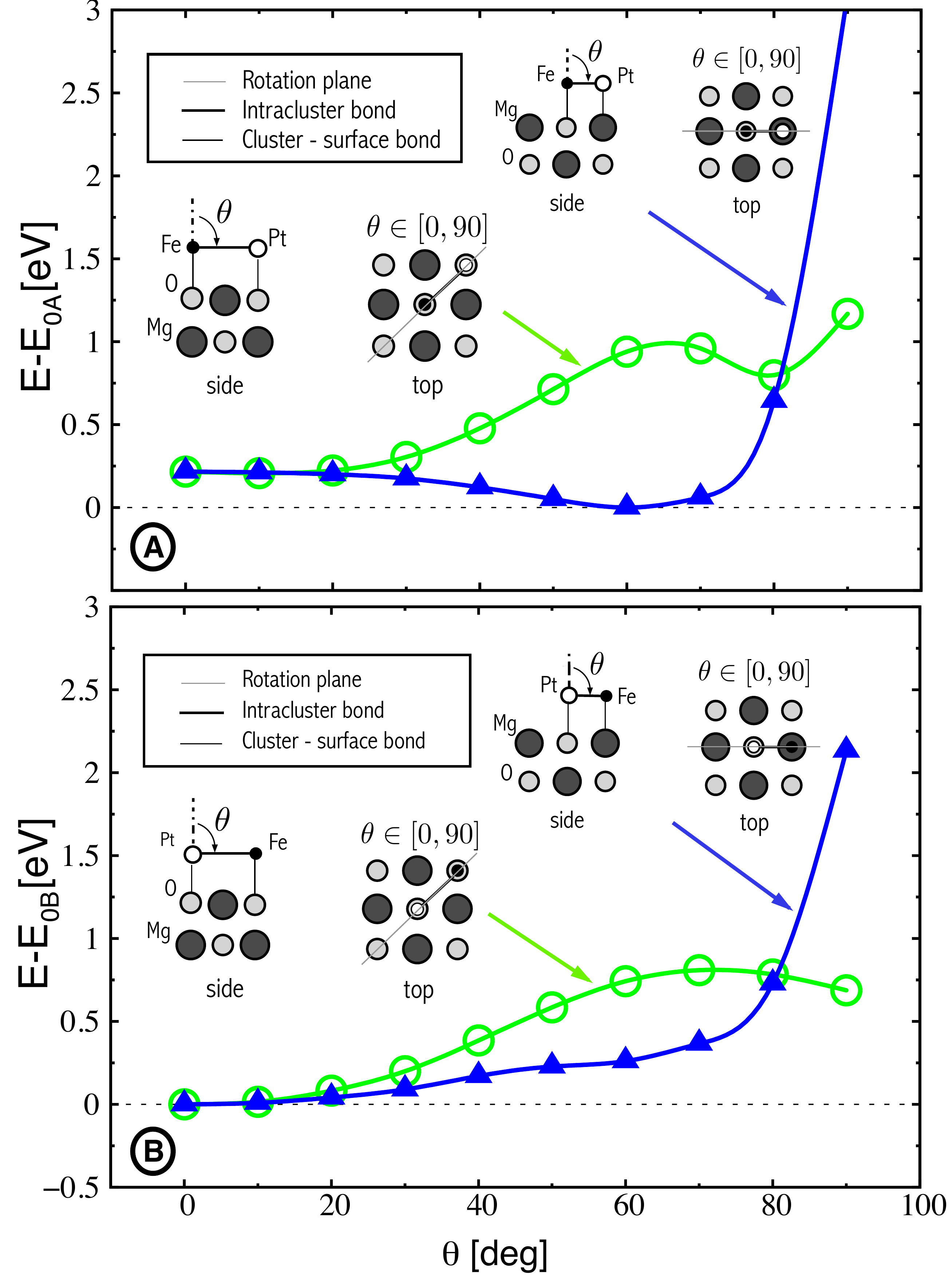}
 \caption{(Color online) Variation of the total energy of an FePt dimer 
          supported on an MgO surface with the angle of the dimer 
          bond, $\theta$ (measured from the normal to the MgO plane). In 
          (A) the Fe side of the dimer is closer to the 
          MgO surface and the dimer is initially placed vertically. The 
          scans take place in two different planes: (011)~(full blue 
          triangles) and on (101)~(empty green circles). In (B) the Pt 
          atoms are placed closer to the surface and the scans are carried 
          out as in (A). In both figures the total energies are plotted 
          relative to the minimum energy for each arrangement~(E$_{0A}$ 
          and E$_{0B}$). The solid lines are a guide for the eye.} 
 \label{dimer-fig}
\end{figure}

Figure~\ref{dimer-fig} summarizes the total energy profile of
the FePt dimer adsorbed onto the MgO surface as a function of the angle 
$\theta$. As the schematic adsorption geometries insets show in the 
figure, the $\theta=0^\circ$ corresponds to the first configuration, 
that is when the dimer lies just out--of--plane with respect to the MgO 
surface; the following total energy values were calculated at $10^\circ$ 
intervals along a circular trajectories on the (011) and (101) rotation 
planes, depending on whether the farthest Fe or Pt atom was conducted to 
lie @$top$--O (green empty circles) or @$top$--Mg~(blue full triangles), 
respectively. This ensures that we checked two possible adsorption sites 
for the outermost atoms. In addition, and based on the convergence test in 
section~\ref{test-subsec}, we calculated two possible adsorption 
configurations depending on whether the Fe or Pt atoms lie @$top$--O, 
Fig.~\ref{dimer-fig}(A) and (B), respectively. Each calculation fixes 
the dimer bond value and only the angle is changed at each step. 
To construct the $\theta=0^\circ$ configuration, the distance between the 
O and Fe or Pt atom was optimized moving away and approaching the dimer 
to the surface. The minimum of the quadratic energy curve gives the Fe/Pt 
bond distance between the atoms and the O site.     

\begin{figure*}
 \includegraphics[scale=0.35]{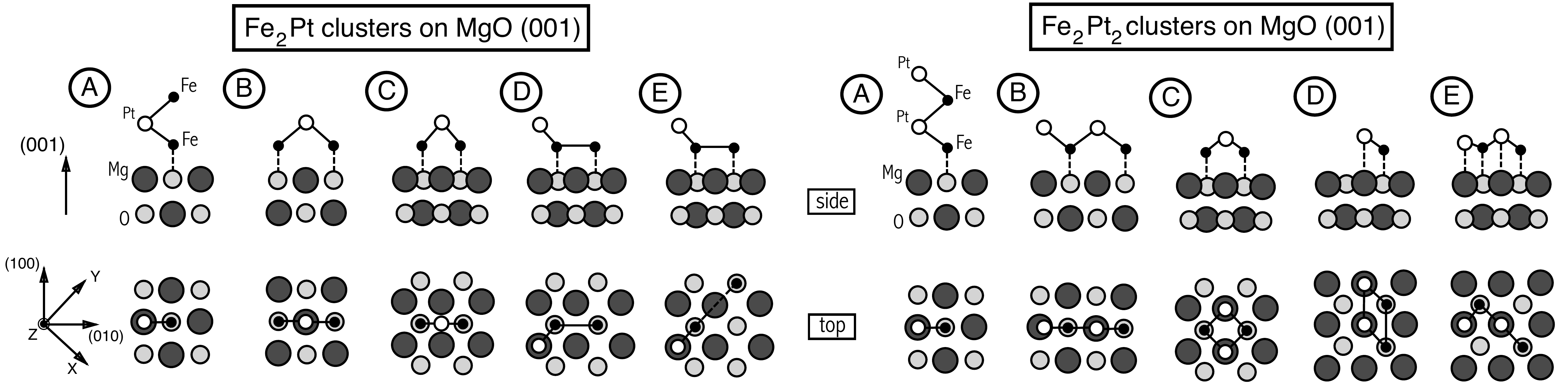}
 \caption{Schematic representation of the initial configurations for the 
          different adsorption positions of the Fe$_2$Pt and Fe$_2$Pt$_2$ 
          clusters on MgO(001) surface, left~(A--E) and right~(A--E), 
          respectively. The side~(top) views are depicted in the 
          upper~(lower) row. In this schematic picture of the adsorption
          of the Fe$_x$Pt$_y$ clusters on MgO surface only the Fe species 
          are directly in contact with the surface via the O atoms. However, 
          along the text, the study of the Pt species on the MgO surface 
          have also taken into account.}  
 \label{tr+te-fig}
\end{figure*}
 
 Consider first the A arrangement, where the Fe is situated on @$top$--O. The
dependence of $E$ on angle is shown for the cases of Pt moving towards Mg and O
sites. The rotation toward the @$top$--O site gives generally a monotonic increase,
albeit with a local minimum close to 90 deg. The global minimum occurs for Pt 
rotating toward the top Mg site at an angle of around 60 deg. In comparison, the 
case B arrangement has a local minimum at 0$^\circ$, i.e with the dimer oriented
perpendicular to the MgO plane. We note that the energies are plotted relative 
to the minimum energy, and that E$_{0A}<$ E$_{0B}$. Thus the global minimum for 
all configurations occurs for the A arrangement with Pt rotated toward the Mg site.

In general, although in this case we only have a dimer on MgO and we are far 
from the FePt bulk phase, we note that the angle is close to the FePt--L1$_0$ 
bulk phase~($\approx46^\circ$). This is a first attempt to discriminate how 
the FePt--L1$_0$ grows onto the MgO alloy. 

\subsubsection{Trimers and tetramers on MgO}\label{tri+tet-subsec}
In the next step we increase the number of atoms in the FePt cluster 
to three, four and so on. Figure~\ref{tr+te-fig} shows the schematic 
configurations of the initial Fe$_2$Pt~(left) and Fe$_2$Pt$_2$~(right) 
adsorbed clusters onto the MgO(001) surface. In both cases we have labeled
them from A to E, though it should be noted that they appear in no particular 
order. In the picture we only show the configurations in which the Fe atoms 
are fixed to the O sites. Those in which the Pt atoms are fixed in the same 
way as the Fe have also been studied, but they are not shown in figure
\ref{tr+te-fig}. The furthest Fe(Pt) atom from the surface in Fe$_2$Pt(Fe$_2$Pt$_2$) 
was placed at several adsorption sites on the surface in order to scan the 
total energy dispersion and enable us to discriminate the preferential growth 
type of the FePt--L1$_0$ onto MgO alloy. 

Among all the configurations, A would represent a vertical growth, whilst the 
others correspond to the FePt atoms wetting the MgO. Within the four B--E 
geometries we will discriminate how far the Fe/Pt species prefer to lie from 
each other while the FePt alloy grows. For instance, in the Fe$_2$Pt B and C 
geometries, the Pt atom is located @$top$--Mg or @$hollow$ site, respectively.
The remaining two settings accommodate the second Fe closer~(D) or farther~(E) 
from the first Fe fixed @$top$--O. Similar situations are depicted for tetramers
(B--E). The starting configurations for the trimer were constructed as follows:
keeping the best dimer geometry, i.e., $60^\circ$ we added one more atom further 
away from the surface in A and at different sites for the remaining places~(B--E). 
It is worth mentioning that, as we showed earlier in~\ref{dimers-subsec}, when 
the Pt atom was closer to the O site the best configuration was vertical, so we 
would expect the lowest energy to be obtained by adding the next Pt to this dimer. 
However, since we want to compare quantitatively  the total energies of similar 
geometries, initially we placed this additional Pt atom at the same position as
the previous one. In the case of the tetramer we had to proceed more carefully
because more initial configurations were possible. In order to reduce them we 
just inspected some. In Fg. \ref{tr+te-fig}(left) configuration (A), we added 
one Fe/Pt, atom following the usual construction of a hypothetical FePt--L1$_0$
unit cell. In the remainder we only placed the Fe/Pt in different sites. Again 
it must be pointed out that additional possibilities will arise when the number 
of atoms increases, but the aim of the present study is to have an initial
indication of how the FePt ``{\it covers}'' the MgO so that a few reasonable
geometries would in principle be enough.  

\begin{figure}[t]
 \includegraphics[scale=0.16]{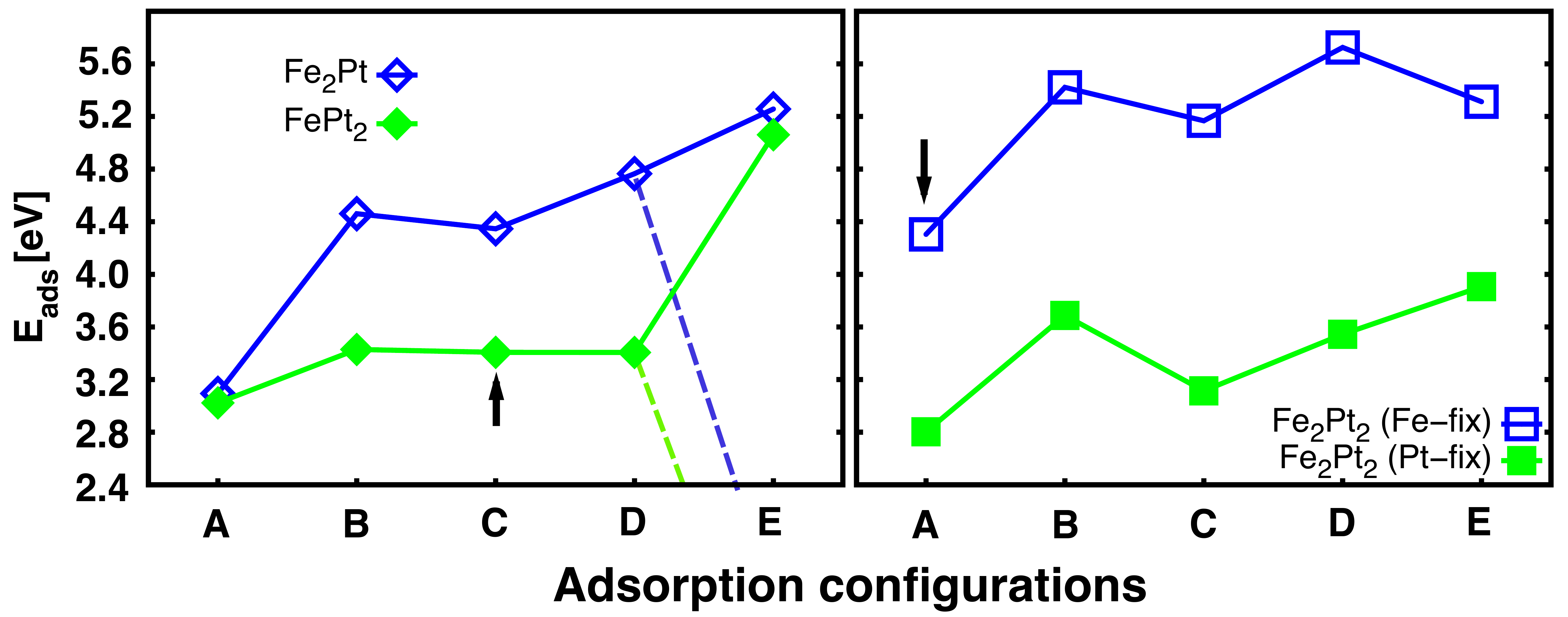}
 \caption{(Color online) Diferent adsorption positions prior to 
          the adsorption energy. {\bf Left}: Fe$_2$Pt and 
          FePt$_2$ $trimers$, blue empty and filled green diamonds, 
          respectively. The blue and green dashed lines represent
          the hypothetical link between the D and E configurations 
          after subtracting the adsorption energy of the isolated Fe 
          or Pt atom; {\bf Righ}t: Fe$_2$Pt$_2$ $tetramers$, blue empty 
          and green filled squares whether the first contact atoms 
          is Fe or Pt, respectively. The black arrows show two 
          configurations that changed substantially after the 
          relaxations.}   
 \label{eads-fig}
\end{figure}

After the relaxation, the shapes of all the initial geometries were kept, just 
some of the atoms in FePt$_2$--C and Fe$_2$Pt$_2$--A~(Fe-terminated) moved away 
from their initial sites. These structures are marked in Fig.~\ref{eads-fig} with 
black arrows. In the first case, the Fe located @$hollow$ initially, moved to lie 
just @$top$--Mg, and in the second, the initial geometry is broken having finally 
each Pt atom @$top$--O/Mg keeping the Fe atoms @$top$--O.  

In principle, the higher the adsorption energy, the more stable the structure. 
In this respect, we observe that the FePt$_2$/Fe$_2$Pt--E geometries  
correspond to the highest E$_{ads}$ and they should be the most stable.
However, we have to take into acount that in the $trimer$--E cases one of the 
Fe/Pt atom is located farther away with respect to the other FePt structure 
and it could be treated as an ``isolated'' atom added to the $dimers$. So, 
as we pointed out in table~\ref{ads-energies}, whether one Fe or Pt atom 
is adsorbed onto a $c32$ supercell using four MgO planes the E$_{ads}$ takes 
the value 2.29~eV for Fe and 3.62~eV for Pt. Roughly, we can subtract these 
values from those in Fig.~\ref{eads-fig} and we would get the values of
$\approx$2.0~eV for FePt$_2$ and $\approx$1.5~eV for Fe$_2$Pt. This immediately 
would indicate that these configurations would not belong to those preferential 
geometries for the FePt on MgO and we can discard them from the preferential 
growth modes. We justify this conclusion by comparing the distances between Fe/Pt 
pairs in E and the other configurations. In E, the Fe--Fe distance is 4.3~\AA\ 
and 4.71~\AA\  for Pt--Pt. The calculated Fe--Fe bond distance for the other 
geometries is around 2.4\AA\ and 2.8\AA\ for the Pt--Pt. Furthermore, if we 
focus on, for example,  the B configuration that has a mediated Fe or Pt atom, 
the distances are 4.1\AA\ and 3.7\AA, for Fe--Fe and Pt--Pt, respectively.
In Fig.~\ref{eads-fig} the blue and green dashed lines represent the connection 
between the D and E adsorption energies in the case that we would assume the 
previous behaviour. It is then interesting to note that the three B, C and D 
geometries have almost a similar adsorption energy values indicating that 
the $trimers$ preferentially spread on MgO instead of growing vertically. As we 
pointed out, the preferential adsortion is when Fe atoms are closer to the 
MgO as the blue line shows. 

The E$_{ads}$ for the $tetramers$ shown in Fig.~\ref{eads-fig}(right) 
tends to corroborate our argument from the previous paragraph, namely, 
that from B up to E geometries the adsorption energies increases 
with respect to A, indicating a predisposition of the Fe/Pt atomic
species to lie onto MgO, covering the surface. It is worth nothing that
the E configuration for $trimers$ is not present for $tetramers$ due to the 
fact that we scanned a limited number of geometries as we pointed out before. 
The blue line in figure~\ref{eads-fig}(right) shows that the $tetramers$ 
follow the same trend as the $trimers$, having the higher adsorption energies 
to those which the Fe atoms are in contact with the Mg surface. We observe in 
B, C, D and E an oscillation in the E$_{ads}$ but its dispersion is only about 
0.5~eV. An additional study of the geometry, as for example the bond and angle 
distances, would explain the origin of this oscilation, but this study is beyond 
the scope of the current work. However, we will compare the cohesive energy of 
the clusters in their gas phases with respect to those adsorbed onto the MgO 
surface in order to check the stability. 

\begin{figure}[b]
 \includegraphics[scale=0.15]{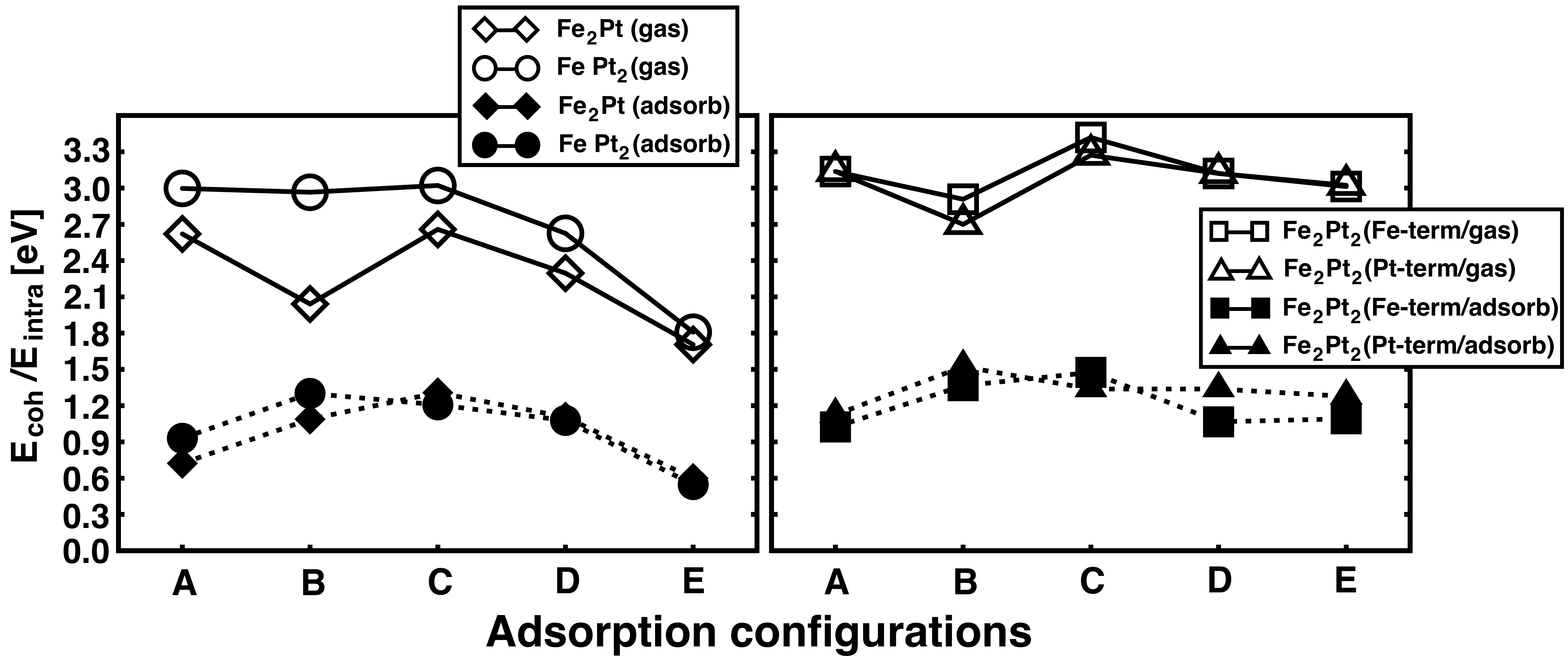}
 \caption{(Color online) Cohesive~(gas phases) and $intracluster$ binding 
          energies~(supported) as a function of the different adsorption 
          positions for the $trimers$ and $tetramers$, left and right graphs, 
          repectively. Empty symbols depict the E$_{coh}$ for the clusters 
          after remove them from the MgO surface and the filled ones the 
          $intracluster$ energy after the adsorption.}   
 \label{cohes-fig}
\end{figure}

In Fig.~\ref{cohes-fig} is shown the cohesive and $intracluster$ binding 
energy of each cluster adsorbed at different $i$ sites before 
and after the adsorption onto the MgO surface, empty and filled symbols, 
respectively. The comparison of these binding values will give an idea 
of the stability of the clusters upon the adsorption and also which of 
them presents more stability at the surface. The binding energies were 
calculated following the equations: 
\[
E_{coh}^{G}(i) = -[E_{G,i}^T - n_{Fe}E_{Fe,i}^{T,G}-n_{Pt}E^{T,G}_{Pt,i}]/N,
\]
for the gas phases~(G), and by: 
\[
E_{intra}^{A}(i)= -[E^T_{A,i} + (N-1)E_{C,i}^T 
           - n_{Fe}E_{Fe,i}^{T,A}-n_{Pt}E^{T,A}_{Pt,i}]/N
\]
after the adsorption~(A). $E_{G,i}^T$ is the total energy of the free 
cluster, $E^T_{A,i}$ is the total energy of the system upon the adsorption, 
$E_{C,i}^T$ is the total energy of the clean MgO surface, $E_{Fe/Pt,i}^{T,G}$ 
is the total energy for an isolated Fe or Pt atom, $E_{Fe/Pt,i}^{T,A}$ is 
the total energy for the surface on which the Fe or Pt atom is adsorbed, 
$n_{Fe,Pt}$ are the number of each specie in the cluster and $N=n_{Fe}+n_{Pt}$ 
is the total number of atoms in the cluster. The values of the $E^{T,A}_{Fe/Pt,i}$ 
were calculated after performing a self consistent~(SC) calculation and keeping 
the closest Fe or Pt atom @$top$--O/--Mg as they were in the final relaxed 
configuration in the $trimers$ and $tetramers$. The total energy of the 
clean surfaces, $E_{C,i}^T$, was calculated by removing the cluster from the 
relaxed structures. 

It is clear that the clusters in their gas phases are more stable than 
those adsorbed onto the MgO surface since the energy has reduced by
an average value of 1.5~eV for the $trimers$ and 2.0~eV for the $tetramers$. 
In this respect the $tetramers$ present a slightly higher stability as indicated 
by the fact that the bigger the cluster, the more stable the structure. 
There is no significant difference in the cohesive and binding energy depending
on whether the Fe or Pt atom is in contact with the surface. In general, the 
most stable configurations after the adsorption, are B and C for $trimers$ and 
$tetramers$. As we pointed out, these geometries would correspond to an initial 
covering of the MgO~(See Fig.~\ref{tr+te-fig}). However it is worth mentioning 
that the competition between the adsorption on the surface and cohesion among the 
atoms in the clusters is an important feature of the Fe$_x$Pt$_y$/MgO system.

\subsection{Charge analysis and Magnetic Moments}\label{dos-sec}
\begin{figure*}[thb]
 \includegraphics[scale=0.20]{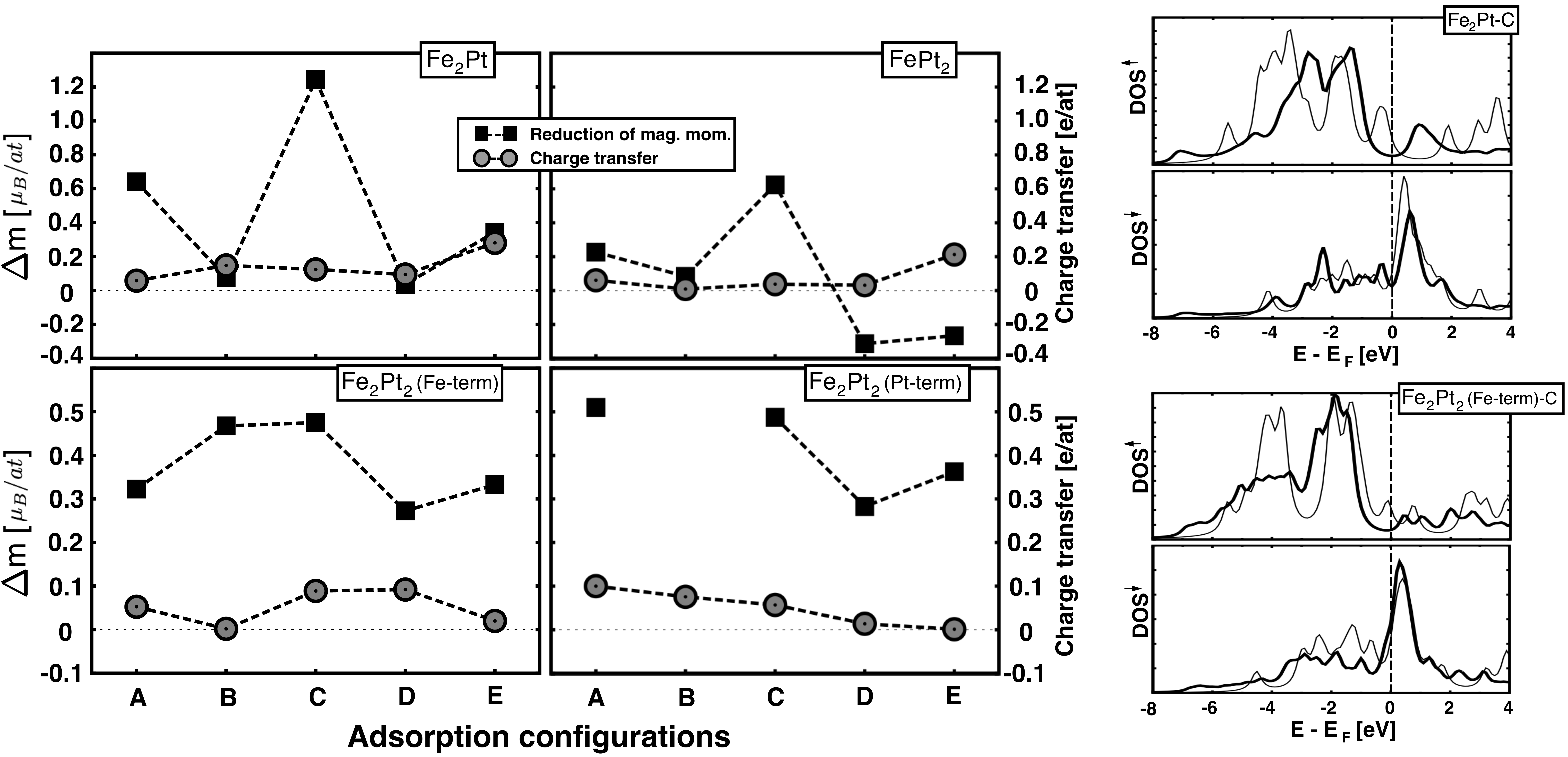}
 \caption{{\bf Left}: Reduction of the magnetic moment, $\Delta MM=MM^G-MM^A$, 
          per atom~(black squares) and charge transfer per atom~(grey circles) 
          as a function of the A to E geometries for $trimers$~(first row) and 
          $tetramers$~(second row). {\bf Right}: Spin-up~(down) projected 
          density of states, DOS$^{\uparrow(\downarrow)}$, for the Fe$_2$Pt-C 
          and Fe$_2$Pt$_2$(Fe-term)-C configurations, upper and lower two
          figures, respectively. Thick solid line represents the projected 
          DOS for the adsorbed clusters and the thin solid lines those in gas
          phases. The DOS units are as usual in states/eV.}
 \label{charge-fig}
\end{figure*}

In this section we address the Mulliken population analysis of each configuration 
to identify any charge transfer between the cluster and the surface atoms. The 
magnetic moment values~(MM) were obtained by subtracting the spin--up from 
spin--down populations. As we will see, upon adsoprtion, the net MM values of 
the $dimers$ and $trimers$ are reduced compared to the gas phases. The mechanisms 
that promote this change have two different origins: geometric and electronic. 
A rearrangement of the atoms in space after the ionic relaxation would imply a 
variation in the population of the cluster's states changing the net MM values. 
However, after comparing the initial and final geometries there were no significant 
changes in the shape of the clusters, so we argue that the geometric disposition 
of the atoms does not influence too much the MM values so that the dominating 
contribution is most likely to be purely electronic. We separate the electronic part 
into two contributions: the first one sets up a charge transfer from the cluster
to the surface, depopulating its states. The second, an $intra$--cluster 
rearrangement of the spin--up and spin--down charges internally increases the up 
or down population at the expense of the other. The gas phases of the adsorbed 
clusters present higher values of their total MM than those upon adsorption. 
Figure~\ref{charge-fig} presents, on the left, the reduction in the MM values~(full 
black squares) for the $trimers$ and $tetramers$~(upper and lower row respectively) 
as a function of the A to E geometries. The total charge transfer per atom from
the clusters to the surface~(grey filled circles) is also shown. On the right we 
have plotted the up and down density of states~(DOS) for the Fe$_2$Pt--C and 
Fe$_2$Pt$_2$(Fe--term)--C geometries as representative examples. As we distinguish 
in the figure, not all the configurations after the adsorption present this
reduction, in particular Fe$_2$Pt--(B,D) and FePt$_2$--(A,B). The reduction will 
be a complex balance due to the $intra$--cluster spin--up/down charge transfer and
also the surface to cluster charge transfer. 

The reduction of the MM is not principally caused by the charge flowing 
between the cluster and surface in all the cases but it is due to the 
$intra$-cluster electron displacements as well. On the right, the DOS$^\uparrow$ 
for the Fe$_2$Pt--C and Fe$_2$Pt$_2$(Fe--term)--C configurations show a hump below
the Fermi level for the gas phases~(grey thin solid line) around $-$0.5~eV for the
first case and at $-$0.2~eV for the second. Upon the adsorption these peaks move 
to the conduction band implying a reduction in the up states. Part of this charge
moves to the O sites and passes internally to the down states. The down states of 
the adsorbed clusters show a slightly different shape in the DOS compared to those 
in gas phases. As we demonstrated the charge balance between the cluster and the
surface is quite complicated, nonetheless our data allow a useful initial attempt 
to explain its behaviour.   

In table~\ref{mm-tab} we summarize the MM/at values of all the configurations
for the magnetic and the nonmagnetic species of the clusters as well as the total
value in bold on the third column of each type of cluster configuration. For the 
$trimers$, the Fe$_2$Pt--$i$ configurations have higher MM values than those 
of the FePt$_2$--$i$ due to the presence of two magnetic atoms instead of only 
one. The Fe$_2$Pt shows the same MM values except the C configuration that has 
0.56~$\mu_B$/at lower. More significant is the change for FePt$_2$ wose MM value, 
depending on the adsorption site,  can vary by up to 0.57~$\mu_B$/at. Coversely 
the trimers present the same MM values for all the adsorpion sites. Upon inspection 
of table~\ref{mm-tab} and figure~\ref{eads-fig} we can conclude that there is no 
strong relationship between the magnetic behaviour and the adsorption energies. 

\begin{table*}[t]
\caption{\label{mm-tab}
          Magnetic moment~(MM) values in $\mu_B$/at of all the clusters
          deposited onto the MgO(001) surface as well as those obtained 
          for the corresponding bulk alloys using the FR-PP formalism 
          The first column displays the adsorption configurations, A--E, as
          well as the bulk. The MM values in the first and the second
          column of each kind of adsorbed geometry have been calculated by
          means of the equations MM$_{M}$/N$_{M}$ and MM$_{NM}$/N$_{NM}$, 
          respectively, whereas in the text, M refers to the magnetic atoms 
          and NM to the non-magnetic ones. The third column shows 
          MM$_{tot}$/N$_{tot}$.}
\begin{tabular}{cccccclccccccclccccccclcccccccl} \hline \hline
 & \multicolumn{6}{c}{Fe$_2$Pt} &&& \multicolumn{6}{c}{FePt$_2$} &&& \multicolumn{6}{c}{Fe$_2$Pt$_2$(Fe-term)} &&& \multicolumn{6}{c}{Fe$_2$Pt$_2$(Pt-term)} \\ \hline
Configuration && M    && NM   &&{\bf Total} &&&& M && NM &&{\bf Total} &&&& M && NM &&{\bf Total} &&&& M && NM && {\bf Total} \\ \hline
{\bf A}       && 3.53 && 0.84 && {\bf 2.63} &&&& 4.09 && 0.54 && {\bf 1.72} &&&& 3.55 && 0.41 && {\bf 1.98} &&&& 3.63 && 0.33 && {\bf 1.98}  \\
{\bf B}       && 3.64 && 0.65 && {\bf 2.64} &&&& 4.12 && 0.82 && {\bf 1.92} &&&& 3.39 && 0.60 && {\bf 2.00} &&&& 3.72 && 0.30 && {\bf 2.01}  \\
{\bf C}       && 2.97 && 0.30 && {\bf 2.08} &&&& 3.77 && 0.16 && {\bf 1.36} &&&& 3.36 && 0.61 && {\bf 1.99} &&&& 3.85 && 0.18 && {\bf 2.01}  \\
{\bf D}       && 3.61 && 0.67 && {\bf 2.63} &&&& 3.85 && 0.53 && {\bf 1.63} &&&& 3.44 && 0.52 && {\bf 1.98} &&&& 3.73 && 0.31 && {\bf 2.02}  \\
{\bf E}       && 3.67 && 0.56 && {\bf 2.63} &&&& 3.83 && 0.11 && {\bf 1.35} &&&& 3.50 && 0.46 && {\bf 1.98} &&&& 3.69 && 0.33 && {\bf 2.01}  \\
{\bf bulk}$^a$&& 3.12 && 0.18 && {\bf 1.65} &&&&  -    &&   -   &&     -    &&&&   -   &&  -    &&     -       &&&&   -   &&  -    &&    -         \\
{\bf Bulk-Others}$^b$ && 2.96 && 0.34 && {\bf 1.65}&&&& - && - && - &&&& - && - && - &&&& - && - &&  -    \\ \hline \hline
\multicolumn{25}{l}{\footnotesize{a~\cite{LS-paper} , b~\cite{galanakis}.}}
\end{tabular}
\end{table*}

\section{Conclusions}\label{conclusion-sec}
We have performed an extensive survey of an ampler set of adsorption positions 
of the Fe$_x$Pt$_y$~($x,y \leq 1,2$) ultrasmall clusters on MgO(001). 
To evaluate the preferential adsorption geometry of FePt on MgO we have calculated 
the energies associated with the adsorption of $dimers$, $trimers$ or $tetramers$.
Furthermore, for each cluster size and shape we kept the Fe or Pt atom @$top$--O 
and moved the outermost ones onto several MgO sites. The schematic configurations 
are depicted in Fig.~\ref{tr+te-fig}. The structural analysis shows that the best 
adsorption geometries among all those studied are those in which they covered the 
surface avoiding  vertical growth, and thereby promoting the  wetting of MgO by 
FePt. After the adsorption, the $intracluster$ energy has decreased by 1.5~eV and 
the the net MM reduces its values for all the configurations at the expense of a
depopulation of the Fe up--$d$ states, transfering part of this charge to the Pt
atoms and externally to the O sites. We acknowledge that the charge balance as well
as the E$_{ads}$  is a complex issue, however our study shows that the overall
conclusion is that the FePt wets the MgO(001) surface. However, we also note that
this is an initial study investigating the fundamental interaction potential between
FePt and MgO. At non--zero temperatures the degree of wetting will be determined by 
a surface free energy, which requires molecular dynamics calculations and is beyond
the scope of the current paper. However, the configurations studied here suggest a
promising starting point for the calculation of the relevant n--body contributions 
to the FePt/MgO interaction potential which the MD calculations will require. 

Finally, the complex charge transfer processes at the FePt--cluster/MgO 
interface predicted here might be expected to be reflected in changes in the 
FePt MAE after the adsorption onto the MgO surface. However, this is an 
interesting possible effect, which is beyond the scope of the present work 
but certainly worthy of further investigation.

\section{Acknowledgments}
The authors are grateful to Dr. Matthew I. J. Probert, Dr YK Takahashi and Dr. T. J. Klemmer for helpful discussions. The financial support of the EU Seventh Framework 
Programme under Grant No. 281043, FEMTOSPIN is gratefully 
acknowledged.


\begin{thebibliography}{References}
\bibitem{shou}
  Sun S., Murray C. B., Weller D., Folks L. and Moser A.
   Science {\bf 287}, 1989 (2000).
\bibitem{hans}
   Hans-Joachim Freund
   Surf. Sci. {\bf 500}, 271 (2002).
\bibitem{musolino}
   V. Musolino, A. Selloni, and R. Car
   Phys. Rev. Lett. {\bf 83}, 3242 (1999).
\bibitem{ricci}
   Davide Ricci, Angelo Bongiorno, Gianfranco Pacchioni, and Uzi Landman
   Phys. Rev. Lett. {\bf 97}, 036106 (2006).
\bibitem{weller}
   D. Weller and A Moser
   IEEE Trans. Mag. {\bf 36}, 10 (2000)
\bibitem{thin}
   J. Kohlhepp and U. Gradmann
   J. Magn. Magn. Mater. {\bf 139}, 347 (1995).
\bibitem{multi}
   J. Lindner, C. R\"udt, E. Kosubek, P. Poulopoulos, K. Baberschke, P. Blomquist, R. W\"appling and D. L. Mills
   Phys. Rev Lett. {\bf 88} ,167206 (2002).
\bibitem{Chepulskii2012}
   Roman V. Chepulskii and W. H. Butler
   Appl. Phys. Lett., {\bf 100}, 142405, (2012).
\bibitem{ramon2}
   R. Cuadrado, Timothy J. Klemmer and R. W. Chantrell
   Appl. Phys. Lett. {\bf 105}, 152406 (2014).
\bibitem{dieguez}
   O. Di\'eguez, M. M. G. Alemany, C. Rey, Pablo Ordej\'on and L. J. Gallego
   Phys. Rev. B {\bf 63}, 205407 (2001).
\bibitem{cuadrado1}
   R. Cuadrado and R. W. Chantrell 
   Phys. Rev. B {\bf 86}, 224415 (2012).
\bibitem{gambardella}
   P. Gambardella, S. Rusponi, M. Veronese, S. S. Dhesi, C. Grazioli, A. Dallmeyer, I. Cabria, R. Zeller, P. H. Dededrich, K. Kern, C. Carbone and H. Brune
   Science {\bf 300}, 1130 (2003).
\bibitem{medina} 
   R. F\'elix-Medina, J. Dorantes-D\'avila and G. M. Pastor
   Phys. Rev. Lett. {\bf 75}, 326 (1995).
\bibitem{rollman} 
   Georg Rollman, Markus E. Gruner, Alfred Hucht, Ralf Meyer, Peter Entel, Murilo L. Tiago and James R. Chelikowsky
   Phys. Rev. Lett. {\bf 99}, 083402 (2007).
\bibitem{seivane} 
   Lucas Fern\'andez--Seivane and Jaime Ferrer
   Phys. Rev. Lett. {\bf 99}, 183401 (2007).
\bibitem{chen}
   K. Chen, S. Fiedler, I. Baev, T. Beeck, W. Wurth and M. Martins
   New Journal of Physics {\bf 14}, 123005 (2012).
\bibitem{bertrand}
   E. Bertrand, D. Bonn, D. Broseta, N. Shahidzadeh, K. Ragil, H. Dobbs, J. O. Indekeu, and J. Meunier 
   J. Pet. Sci. Eng. {\bf 33}, 217 (2002).
\bibitem{tabeling}
   P. Tabeling
   Microfluidics (EDP Sciences, Paris) (2004).
\bibitem{siesta}
   J.M. Soler, E. Artacho, J.D. Gale, A. Garc\'ia, J. Junquera, P. Ordej\'on and D. S\'anchez-Portal,
   J. Phys.: Condens. Matter, {\bf 14}, 2745, (2002).
\bibitem{pbe}
   J. P. Perdew, K. Burke and M. Ernzerhof,
   Phys. Rev. Lett., {\bf 77}, 3865, (1996).
\bibitem{kb}
   L. Kleinman and D. M. Bylander,
   Phys. Rev. Lett., {\bf 48}, 1425, (1982).
\bibitem{tm}
   N. Troullier and J. L. Martins,
   Phys. Rev. B, {\bf 43}, 1993, (1991).
\bibitem{cc}
   S.G. Louie, S. Froyen and M.L. Cohen,
   Phys. Rev. B, {\bf 26}, 1738 (1982).
\bibitem{FePtOnMgO}
   R. Cuadrado and R. W. Chantrell
   Phys. Rev. B {\bf 89}, 094407 (2014).
\bibitem{LS-paper}
   R. Cuadrado and J. I. Cerdá
   J. Phys.: Condens. Matter {\bf 24}, 086005 (2012).
\bibitem{galanakis}
   I. Galanakis, M. Alouani, H. Dreysse,
   Phys. Rev. B, {\bf 62}, 6475, (2000).
\end{thebibliography}
\end{document}